\documentclass[10pt]{iopart}
\usepackage{iopams}
\begin{document}
\jl{8}
\def\phi{\varphi} \def\epsilon{\varepsilon} \def\u{\boldsymbol}


\title{Scaling law for the critical function of an approximate
  renormalization}

\author{C Chandre\footnote[1]{E-mail:
  \texttt{chandre@spht.saclay.cea.fr}} and P
  Moussa\footnote[2]{E-mail: \texttt{moussa@spht.saclay.cea.fr}} }
\address{Service de Physique
  Th\'eorique, CEA Saclay, F-91191 Gif-sur-Yvette Cedex, France}
\date{\today}

\begin{abstract}
  We construct an approximate renormalization for Hamiltonian systems
  with two degrees of freedom in order to study the break-up of
  invariant tori with arbitrary frequency. We derive the equation of  
  the critical surface of
  the renormalization map, and we compute the scaling behavior of the
  critical function of one-parameter families of Hamiltonians, near
  rational frequencies. For the forced pendulum model, we find the
  same scaling law found for the standard map in [Carletti and Laskar,
  preprint (2000)]. We discuss a conjecture on the link between the
  critical function of various types of forced pendulum models, with the
  Bruno function.
\end{abstract}
\pacs{05.45.Ac, 05.10.Cc, 45.20.Jj}
\submitted 

\maketitle

\section{Introduction}
\label{sec:intro}

In this article, we analyze the critical function of one parameter
families of Hamiltonians near a resonance, by an approximate
renormalization group method. We consider the following Hamiltonians
with two degrees of freedom written in actions $\u{A}=(A_1,A_2)\in
{\Bbb R}^2$ and angles $\u{\phi}=(\phi_1,\phi_2)\in {\Bbb T}^2$~:
\begin{equation}
  \label{eq:hamg}
  H(\u{A},\u{\phi})=H_0(\u{A})+\varepsilon V(\u{\phi}).
\end{equation}
where the integrable part $H_0$ of the Hamiltonian is
\begin{equation}
  \label{eq:h0}
  H_0(\u{A})=\u{\omega}\cdot\u{A}+\frac{1}{2}(\u{\Omega}\cdot\u{A})^2.
\end{equation}
The vector $\u{\omega}=(\omega, -1)$ is the frequency vector of the
considered torus which is located at $\u{\Omega}\cdot\u{A}=0$ for
$H_0$. For $\varepsilon$ sufficiently small and for $V$ sufficiently
smooth, the KAM theorem ensures the persistence of this
torus for a diophantine frequency
$\omega$~\cite{kolm54,arno63a,mose62} (see Ref.~\cite{chan98c} for a
proof in the case of Hamiltonians~(\ref{eq:hamg})-(\ref{eq:h0})). 
When we increase the parameter up to a critical
value, the torus breaks up.  The critical function $\omega\mapsto
\varepsilon_c(\omega)$ is the value of the parameter $\varepsilon$ for
which the invariant torus with frequency $\omega$ breaks up. We notice
that this function vanishes at all rational values of $\omega$. For
$\omega$ close to $P/Q$ satisfying a diophantine condition, 
Carletti and Laskar~\cite{carl00} found numerically,
using the frequency map analysis, the following scaling law for the
standard map critical function~:
\begin{equation}
  \label{eq:sclaw}
  \log \varepsilon_c(\omega) \approx \alpha_{P,Q} \log |\omega-P/Q|
  +c_{P,Q} \qquad \mbox{ for } \omega \approx P/Q,
\end{equation}
with $\alpha_{P,Q}=1/Q$. For the main resonances $0/1$ and $1/1$, they
proved analytically that this exponent is equal to one.\\
We derive an approximate renormalization transformation
${\mathcal{R}}: H \mapsto H'$ acting in some space of Hamiltonians, in
order to analyze the scaling law for different one parameter families
of Hamiltonians~(\ref{eq:hamg}). This transformation gives an
approximate value for the critical function $\varepsilon_c(\omega)$.
The approximate renormalization scheme we use is similar to the one
set up by Escande and Doveil~\cite{esca81,esca85} (see also
Refs.~\cite{mack88,chan99a}). Complete versions of these
transformations have been defined and studied numerically in
Refs.~\cite{koch99,govi97,chan98a,chan98b,abad98,chan00a}. All these
transformations are composed by a sequence of canonical changes of
coordinates~: a partial elimination of the Fourier modes of the
Hamiltonian and a rescaling procedure (shift of the Fourier modes,
rescaling of time and of the actions). The aim of this renormalization
is to change the scale of phase space such that the renormalized
system describes the original one on a smaller scale in phase space
and longer time scale.\\
In the space of Hamiltonians there are two main domains for the
renormalization map~: a domain where the iterations of ${\mathcal{R}}$
converge to the integrable Hamiltonian $H_0$, and another domain where
the iterations diverge to infinity. The two domains are separated by a
surface ${\mathcal{S}}$. We assume that ${\mathcal{S}}$ is a good
approximation of the critical surface which is the set of Hamiltonians
that have a torus at the threshold of the break-up (see
Ref.~\cite{chan99a} for numerical evidence). In what follows, we
will identify ${\mathcal{S}}$ with the critical surface.\\
The approximations we perform consist in mainly two approximations~:
we project 
the transformation on the two main Fourier modes of the perturbation
and we neglect higher orders in $\varepsilon$ (the amplitude of the
perturbation). These approximations reduce the renormalization 
transformation
${\mathcal{R}}$ to a 4 dimensional map from which we can derive
analytical results.\\ 
Following Ref~\cite{mack88}, we derive analytically the equation of  
the critical surface $\mathcal{S}$ for the
approximate renormalization transformation, and we analyze the precise
shape of this surface near a resonance $P/Q$, i.e., when the frequency
$\omega$ is close to $P/Q$.  The result we found is that the scaling
law~(\ref{eq:sclaw}) is satisfied. The 
characteristic exponent $\alpha_{P,Q}$ of the scaling
law~(\ref{eq:sclaw}) depends on 
the one parameter family we consider. For the following
forced pendulum model (Escande-Doveil Hamiltonian~\cite{esca81,esca85})
\begin{equation}
  \label{eq:edfp}
  H(p,x,t)=p^2/2-\varepsilon\left(\cos x +\cos(x-t)\right),
\end{equation}
which shares the same symmetries as the standard map, the 
characteristic exponent is the same as the one for the standard map
critical function~: $\alpha_{P,Q}=1/Q$ for $0<P<Q$. However, this
exponent $\alpha_{P,Q}$ is not the same for all one-parameter family
critical function since we will
exhibit examples of Hamiltonians with different values (see
Eq.~(\ref{eq:chalp}) below). \\

In Sec.~\ref{sec:equ}, we derive the equivalence between the critical
function of a simple Hamiltonian model and the critical function of
the following forced pendulum models~:
$$
H(p,x,t)=\frac{1}{2}p^2 -\varepsilon(\cos(ax-bt)+\cos(cx-dt)),
$$
where $(a,b,c,d)\in {\mathbb{Z}}^4$, $a,c>0$ and $ad-bc=1$.
In Sec.~\ref{sec:def}, we construct the renormalization transformation
by defining the two generators ${\mathcal{R}}_C$ and ${\mathcal{R}}_S$
of this transformation. We give the explicit formulas of this
transformation as a function of the continued fraction expansion of
the frequency of the torus. In Sec.~\ref{sec:cri}, we derive the
equation of the critical surface and we analyze its shape near a
resonance by computing the characteristic exponents of the scaling
laws of the critical function. In Sec~\ref{sec:bru}, we analyze the
scaling law for the forced pendulum models
in order to see the generality of the scaling law and its
characteristic exponent. We also discuss a conjecture on the link
between the critical function and the Brjuno function.

\section{Critical function of various types of forced pendulum
  models} 
\label{sec:equ}

We consider the following families of Hamiltonian systems~:
\begin{equation}
  \label{eq:gfp}
  H=\frac{1}{2}p^2 -\varepsilon_1 \cos(a x-bt) 
  - \varepsilon_2 \cos(c x -d t) ,
\end{equation}
with $(a,b,c,d)\in {\mathbb{Z}}^4$, $a,c>0$ and $ad-bc=1$.  
In particular, we notice that for
$a=c=d=1$ and $b=0$, the Hamiltonian~(\ref{eq:gfp}) with $\varepsilon_1=\varepsilon_2=\varepsilon$ is the forced
pendulum~(\ref{eq:edfp}). For these models, the invariant torus
with frequency $\omega$ is located at $p=\omega$ for the unperturbed
motion $\varepsilon_1=\varepsilon_2=0$. It is equivalent to study the torus located at
$p=0$ of the following Hamiltonian~:
$$
H=\frac{1}{2}p^2+\omega p -\varepsilon_1 \cos(a x-bt) -\varepsilon_2 
\cos(c x -d t) .
$$
This Hamiltonian is also mapped into a time-independent two degrees
of freedom Hamiltonian~:
\begin{equation}
\label{eq:gfp2d}
H(\u{A},\u{\phi})=\u{\omega}\cdot\u{A}+\frac{1}{2}A_1^2-\varepsilon_1
 \cos(a\phi_1+b\phi_2)-\varepsilon_2 \cos(c\phi_1+d\phi_2),
\end{equation}
where $\u{\omega}=(\omega,-1)$. This Hamiltonian belongs to the
family~(\ref{eq:hamg}), and 
it is canonically conjugated with the following Hamiltonian~:
\begin{equation}
  \label{eq:monH}
  \hat{H}(\u{A},\u{\phi})=\u{\omega}\cdot\u{A}+\frac{1}{2}(\u{\Omega}
  \cdot 
  \u{A})^2- f_1 \cos\phi_1-f_2 \cos\phi_2,
\end{equation}
where $\u{\omega}=(\omega,-1)$ and $\u{\Omega}=(1,k)$ for some $k>0$.
The exact generalized canonical transformation which links the two
models is given by three successive steps. The first step is
a shift of the Fourier modes
$(\u{A}',\u{\phi}')=(\tilde{T}^{-1}\u{A}, T\u{\phi})$, where
$T=\left(\begin{array}{cc} a & b\\ c & d\end{array}\right)$ and
its transposed matrix $\tilde{T}$ have
integer coefficients and determinant 1 in order to preserve the
periodic structure of the angles. After the shift of the
Fourier modes, the Hamiltonian~(\ref{eq:gfp2d}) becomes~:
$$
H'(\u{A},\u{\phi})=(d-c\omega)\u{\omega}'\cdot\u{A}+\frac{a^2}{2}(\u{\Omega}'\cdot 
\u{A})^2 -\varepsilon_1 \cos\phi_1 -\varepsilon_2 \cos\phi_2,
$$
where $\u{\omega}'=((a\omega-b)/(d-c\omega),-1)$ and
$\u{\Omega}'=(1, c/a)$. The second step is a
rescaling of time by a factor $1/(d-c\omega)$~: we multiply globally  
the Hamiltonian by a
factor $1/(d-c\omega)$ such that the linear term
becomes $\u{\omega}'\cdot \u{A}$. The third step is a rescaling of the
actions by a factor $\lambda=a^2/(d-c\omega)$~: 
we replace the Hamiltonian $H'$ by $\lambda
H'(\u{A}/\lambda,\u{\phi})$ (this transformation does not change the
equations of motion), with $\lambda=a^2/(d-c\omega)$ such that the
quadratic term is equal to $(\u{\Omega}'\cdot\u{A})^2/2$.
Thus the image of the
Hamiltonian $H$ given by Eq.~(\ref{eq:gfp2d}) by this transformation
is the following one~:
\begin{equation}
\label{eq:eqHH}
H''(\u{A},\u{\phi})=\u{\omega}'\cdot\u{A}+\frac{1}{2}(\u{\Omega}'\cdot 
\u{A})^2 -\frac{a^2}{(d-c\omega)^2}
(\varepsilon_1 \cos\phi_1+\varepsilon_2 \cos\phi_2),
\end{equation}
which is of the same form as $\hat{H}$ with $\omega'=(a\omega-b)/
(d-c\omega )$, $k=c/a$, $f_1=a^2
\varepsilon_1/(d-c \omega)^2$ and
$f_2=a^2\varepsilon_2/(d-c\omega)^2$.\\
We consider the one-parameter family of Hamiltonians~(\ref{eq:gfp})
with $\varepsilon_1=\varepsilon_2=\varepsilon$ which is denoted $F_1$,
and the one-parameter family of Hamiltonians~(\ref{eq:monH}) with
$f_1=f_2=\varepsilon$, denoted $F_2$.
We denote $\varepsilon_{fp}(\omega)$ the critical function of the  
one-parameter family $F_1$, and
$\varepsilon_H(\omega)$ the one of the family $F_2$.  
According to the equivalence between
Hamiltonians~(\ref{eq:gfp}) and Hamiltonians~(\ref{eq:monH}), the  
critical functions are linked by the following law~:
\begin{equation}
\label{eq:numb}
\varepsilon_{fp}(\omega)=a^{-2}(d-c\omega)^2\varepsilon_{H}(\omega'),
\end{equation}
for $\omega\in [b/a,d/c]$, where $\omega'\geq 0$ is given by
\begin{equation}
\label{eq:omom}
\omega'=-\frac{a\omega-b}{c\omega-d}.
\end{equation}
Consequently, we emphasize the role of modular transformations in the
study of critical functions. One already knows that the classical
diophantine conditions are invariant under modular
transformations. Furthermore, the modular properties have been
previously considered in the modular smoothing approach~\cite{buri90} 
which analyzes
the compensations between the singular behaviors at $\omega$ and
$1/\omega$.    

\section{Renormalization operators}
\label{sec:def}

In this section, we derive the approximate renormalization map for a
given irrational frequency $\omega>0$ by defining two elementary
operators ${\mathcal{R}}_C$ and ${\mathcal{R}}_S$. This construction
follows the approximate renormalization developed by MacKay in
Ref.~\cite{mack88}. However the resulting renormalization
transformation is different since in Ref.~\cite{mack88}, the scheme is
based on the analysis of stability of nearby periodic orbits by
looking at the residues.\\
We perform
${\mathcal{R}}_C$ if $\omega>1$ and ${\mathcal{R}}_S$ if
$\omega<1$. The renormalization operator ${\mathcal{R}}_C$ 
is exact (it is generalized
canonical transformation without any truncature), and
${\mathcal{R}}_S$ contains a projection on some relevant Fourier modes
of the Hamiltonian, and we neglect the high orders in the perturbation
parameter.\\
Each of these two elementary renormalization operators acts on the
following family of Hamiltonians~:
\begin{equation}
  \label{eq:hamrp}
  H(\u{A},\u{\phi})=\u{\omega}\cdot\u{A}+\frac{1}{2}(\u{\Omega} \cdot 
  \u{A})^2-f_1\cos\phi_1-f_2\cos\phi_2,
\end{equation}
which belong to the family of Hamiltonians~(\ref{eq:hamg}).
The vector $\u{\omega}=(\omega,-1)$ is the frequency vector of the
considered torus, and the vector $\u{\Omega}=(1,k)$ is the direction
of twist in the angles. We notice that if $kf_1f_2=0$, the
Hamiltonian (\ref{eq:hamrp}) is integrable.

\subsection{Definition of ${\mathcal{R}}_C$}
\label{sec:RP}

The renormalization operator ${\mathcal{R}}_C$ contains a shift of the
Fourier modes and a rescaling of time and of the actions. The shift of
the Fourier modes is constructed such that it exchanges the mode
$(1,0)$ and $(0,1)$. More precisely, we require that the new angles
$\u{\phi}'$ satisfy~:
\begin{eqnarray*}
  && \cos\phi'_1=\cos\phi_2,\\
  && \cos\phi'_2=\cos\phi_1.
\end{eqnarray*}
This is performed by the following linear canonical transformation~:
$$
(\u{A},\u{\phi})\mapsto (\u{A}',\u{\phi}')=(C\u{A},C\u{\phi}),
$$
where $C$ is the following symmetric orthogonal matrix~:
$$
C=\left(\begin{array}{cc} 0 & 1\\ 1 & 0 \end{array}\right).
$$
The vectors $\u{\omega}$ and $\u{\Omega}$ are changed into
$C\u{\omega}=(-1,\omega)$ and $C\u{\Omega}=(k,1)$. We impose that the
images $\u{\omega}'$ and $\u{\Omega}'$ of the vectors $\u{\omega}$ and
$\u{\Omega}$ by the renormalization ${\mathcal{R}}_C$ satisfy the
following normalization conditions~: The second component of
$\u{\omega}'$ must be equal to -1, and the first component of
$\u{\Omega}'$ must be equal to 1.  We rescale the time by a factor
$\omega$, i.e.\ we multiply the Hamiltonian by $1/\omega$, and we
change the sign of phase space coordinates
$(\u{A}',\u{\phi}')=(-\u{A},-\u{\phi})$.  Then $\u{\omega}'$ is
given by $\u{\omega}'=(1/\omega,-1)$, and $\u{\Omega}'=(1,1/k)$.\\
The quadratic part of the Hamiltonian becomes
$k^2(\u{\Omega}'\cdot\u{A})^2/(2\omega)$.  In order that this
quadratic term is equal to $(\u{\Omega}'\cdot\u{A})^2/2$, we rescale
the actions by a factor $\lambda_C=k^2/\omega$, i.e.\ we change the
Hamiltonian $H$ into $\lambda_C H(\u{A}/\lambda_C,\u{\phi})$.\\
In summary, a Hamiltonian $H$ given by Eq.~(\ref{eq:hamrp}) is mapped
into
$$
H'(\u{A},\u{\phi})=\u{\omega}'\cdot\u{A}+\frac{1}{2}(\u{\Omega}'
\cdot \u{A})^2-f_1'\cos\phi_1-f_2'\cos\phi_2,
$$
where $\omega'=1/\omega$, $k'=1/k$, and
\begin{eqnarray*}
  && f_1'=f_2 k^2/\omega^2,\\
  && f_2'=f_1 k^2/\omega^2.
\end{eqnarray*}
The renormalization operator ${\mathcal{R}}_C$ is equivalent to the
4-dimensional map given by the above equations~:
$$
(f_1,f_2,\omega,k)\mapsto(f_1',f_2',1/\omega,1/k).
$$
In order to simplify the computations, we denote
$\tilde{f_2}=k^2f_2/\omega^2$. The renormalization operators is given
by~:
\begin{eqnarray*}
  && f_1'=\tilde{f_2},\\
  && \tilde{f_2'}=f_1.
\end{eqnarray*}
We notice that the new frequency $\omega'=1/\omega$ is smaller than one.

\subsection{Definition of ${\mathcal{R}}_S$}
\label{sec:RL}

The renormalization operator ${\mathcal{R}}_S$ acts on the family of
Hamiltonians~(\ref{eq:hamrp}) for $\omega <1$. 
It contains an elimination of the mode
$(0,1)$ of the perturbation, and a rescaling procedure (shift of the
resonances, rescaling of time and of the actions) such that the image
of a Hamiltonian $H$ given by Eq.~(\ref{eq:hamrp}) is of the same
general form as $H$ and describes the system on a smaller scale in
phase space and on a longer time scale.\\
We eliminate the mode $(0,1)$ of the perturbation, by a near-identity
canonical transformation. We perform a Lie transformation generated by
a function $F(\u{A},\u{\phi})$. The image of a Hamiltonian $H$ is
given by~:
\begin{equation}
  H'=\exp(\hat{F})H=H+\{F,H\}+\frac{1}{2}\{F,\{F,H\}\}+\cdots,
\end{equation}
where $\{\, , \, \}$ denotes the Poisson bracket~:
$$
\{f,g\}=\frac{\partial f}{\partial \u{\phi}}\cdot \frac{\partial
  g}{\partial \u{A}} - \frac{\partial g}{\partial \u{\phi}}\cdot
\frac{\partial f}{\partial \u{A}},
$$
and the operator $\hat{F}$ acts on $H$ like $\hat{F}H=\{ F,H\}$.
The generating function is chosen linear in the actions and of the
form~:
\begin{equation}
  \label{eq:S1}
  F(\u{A},\u{\phi})=-f_2(1+k \u{\Omega}\cdot\u{A})\sin\phi_2.
\end{equation}
The generating function $F$ satisfies~:
$$
\{ F,H_0\}=f_2\cos\phi_2+O(\varepsilon \u{A}^2),
$$
where $\varepsilon$ denotes the order of $f_1$ and $f_2$, and $H_0$
is the integrable part of $H$ given by Eq.~(\ref{eq:h0}).  The image
of a Hamiltonian $H$ given by Eq.~(\ref{eq:hamrp}) is~:
\begin{eqnarray*}
  H'=&& H_0-f_1 \cos\phi_1-f_2 \cos\phi_2+\{F,H_0\}\\
  && -f_1\{F,\cos\phi_1 \}
   -f_2\{F,\cos\phi_2\}+\frac{1}{2}\{F,\{F,H_0\}\}+O(\varepsilon^3).
\end{eqnarray*}
We neglect the order $O(\varepsilon \u{A}^2)$ produced by the
transformation together with the orders $O(\varepsilon^3)$. These
approximations are consistent since for small $\varepsilon$,
$\u{A}(t)$ is of order $\varepsilon$ near the torus with frequency $\omega$. The term
$-f_2\{F,\cos\phi_2\}$ is independent of the actions and contains the
modes $\u{0}$ and $(0,2)$ that we neglect. The term
$-f_1\{F,\cos\phi_1\}$ generates the modes $(1,1)$ and (1,-1). We
neglect the last one and we keep the next relevant Fourier mode
$-\frac{1}{2}k f_1 f_2 \cos(\phi_1+\phi_2)$.  Then after the
elimination of the mode $(0,1)$, the new Hamiltonian is equal to~:
\begin{equation}
  H'=\u{\omega}\cdot \u{A}+ (\u{\Omega}\cdot\u{A})^2/2-f_1\cos\phi_1
 -\frac{1}{2} k f_1 f_2 \cos(\phi_1+\phi_2). \label{eq:h'}
\end{equation}
In this elimination procedure, we have neglected the Fourier modes
$(1,-1)$ and $(0,2)$ which are of the same order $\varepsilon^2$ as
the mode $(1,1)$ that we kept. The justification for this choice is
that the mode $(1,1)$ is associated with the smallest small
denominator $\u{\omega}\cdot\u{k}$ among these three modes~:
$|\u{\omega}\cdot (1,1)|=1-\omega< |\u{\omega}\cdot
(1,-1)|=1+\omega<|\u{\omega}\cdot (0,2)|=2$. Thus we expect that the
dynamics is mainly influenced by the mode $(1,1)$ at the order
$\varepsilon^2$ of the perturbation. In phase space, this means that
the periodic orbit with frequency vector $(1,1)$ is closer to the
torus (with frequency $\omega$) than the periodic orbits with
frequency vectors $(1,-1)$ and $(0,2)$.\\
We shift the Fourier modes according to the following linear canonical
transformation~:
$$
(\u{A},\u{\phi})\mapsto (\u{A}',\u{\phi}')=(S^{-1}\u{A},S\u{\phi}),
$$
where
$$
S=\left( \begin{array}{cc} 1 & 1 \\ 1 & 0 \end{array} \right),
$$
such that $\cos\phi_1=\cos\phi'_2$ and
$\cos(\phi_1+\phi_2)=\cos\phi'_1$. The vectors $\u{\omega}$ and
$\u{\Omega}$ are changed into $S\u{\omega}=(\omega-1,\omega)$ and
$S\u{\Omega}=(k+1,1)$. Thus the normalization conditions on these
vectors give the image of $\omega$ and $k$~:
\begin{eqnarray*}
  && \omega'=(1/\omega)-1,\\
  && k'=1/(k+1).
\end{eqnarray*}
We rescale time in such a way that the linear term in the actions of
$H_0$ is equal to $\u{\omega}'\cdot\u{A}$ with
$\u{\omega}'=(\omega',-1)$~: we multiply the Hamiltonian $H'$ by a
factor $1/\omega$. The quadratic term of the new Hamiltonian is equal
to $ (k+1)^2 (\u{\Omega}'\cdot\u{A})^2/(2\omega)$. In order to map
this quadratic part into $(\u{\Omega}'\cdot\u{A})^2/2$, we rescale the
actions by a factor
$$
\lambda_S= (k+1) ^2/\omega,
$$
by changing the Hamiltonian $H'$ into $\lambda_S
H'(\u{A}/\lambda_S,\u{\phi})$.\\
After this elimination and rescaling procedures, a Hamiltonian $H$ is
mapped into
$$
H''=\u{\omega}'\cdot\u{A}+\frac{1}{2}(\u{\Omega}'\cdot\u{A})^2
-f_1'\cos\phi_1-f_2'\cos\phi_2,
$$
where
\begin{eqnarray*}
  && f_1'=\frac{k(k+1)^2}{2\omega^2} f_1 f_2,\\
  && f_2'=\frac{(k+1)^2}{\omega^2} f_1.
\end{eqnarray*}
The renormalization operator $\mathcal{R}_S$ is equivalent to the
4-dimensional map
$$
(f_1,f_2,\omega,k)\mapsto (f_1',f_2',1/\omega-1,1/(k+1)),
$$
given by the above equations. In the variables
$(f_1,\tilde{f_2}=k^2f_2/\omega^2)$
the renormalization operator is given by~:
\begin{eqnarray*}
  && f_1'=\frac{(k+1)^2}{2k} f_1\tilde{f_2},\\
  && \tilde{f_2'}=\frac{1}{(1-\omega)^2} f_1.
\end{eqnarray*}

\subsection{Renormalization operator ${\mathcal{R}}$}

We define the renormalization operator ${\mathcal{R}}$ according to
the value of the frequency $\omega$ in the following way~:
\begin{eqnarray*}
  && {\mathcal{R}}={\mathcal{R}}_C \quad \mbox{ if } \omega > 1,\\
  && {\mathcal{R}}={\mathcal{R}}_S \quad \mbox{ if } \omega < 1.
\end{eqnarray*}
In order to know the sequence of operators ${\mathcal{R}}_C$ and
${\mathcal{R}}_S$, we look at the continued fraction expansion of the
frequency $\omega$. We denote $a_i$ the entries in the continued
fraction of $\omega$~:
$$
\omega=[a_0,a_1,\ldots]=\frac{1}{a_0+\frac{1}{a_1+\cdot}}.
$$
The sequence of renormalization operators is
$({\mathcal{R}}_S{\mathcal{R}}_C)^{a_0-1}{\mathcal{R}}_S
({\mathcal{R}}_S{\mathcal{R}}_C)^{a_1-1}{\mathcal{R}}_S \cdots$. We
denote ${\mathcal{R}}_{a} =
({\mathcal{R}}_S{\mathcal{R}}_C)^{a-1}{\mathcal{R}}_S$. This
renormalization operator changes the frequency
$\omega=[a_0,a_1,\ldots]$ into $\omega'=[a_1,a_2,\ldots]$, and
$k=[b_0,b_1,\ldots]$ into $k'=[a_0,b_0,b_1,\ldots]$. The explicit
formulas for ${\mathcal{R}}_a$ are the following ones~:
\begin{eqnarray*}
  && f_1'=\rho_1(\omega,k) f_1^a \tilde{f_2},\\
  && \tilde{f_2'}=\rho_2(\omega) f_1,\\
  && \omega'=1/\omega -a,\\
  && k'=1/(k+a),
\end{eqnarray*}
where the coefficients $\rho_1$ and $\rho_2$ are given by
\begin{eqnarray}
  \label{eq:rho1}
  && \rho_1(\omega,k)=2^{-a}(1+a/k)\prod_{l=0}^{a-1}
    (k+l+1)(1-l\omega)^{-2},\\
  && \rho_2(\omega)=(1-a\omega)^{-2}. \label{eq:rho2}
\end{eqnarray}
For $a=1$, we check that $\rho_1(\omega,k)$ and $\rho_2(\omega)$ are
equal to the coefficients of ${\mathcal{R}}_S$.
The relations~(\ref{eq:rho1})-(\ref{eq:rho2}) can be proven recursively, 
using the fact that
${\mathcal{R}}_{a+1}={\mathcal{R}}_S {\mathcal{R}}_C
{\mathcal{R}}_a$.\\
Moreover, the rescaling in the actions after the renormalization
${\mathcal{R}}_a$ is equal to $\lambda_a=(k+a)^2/\omega$, and the
rescaling in time is equal to $1/\omega$. We notice that
${\mathcal{R}}_a$ has two fixed points with
$k=\omega=(\sqrt{a^2+4}-a)/2$~: a trivial one $f_1=\tilde{f_2}=0$, and
a non-trivial one $f_1=(\rho_1\rho_2)^{-1/a}$,
$\tilde{f_2}=\rho_2(\rho_1\rho_2)^{-1/a}$. \\

\textit{Remark~:} In the construction of ${\mathcal{R}}_a$, we have
neglected several terms, some of them are of order
$O(\varepsilon^3)$. Therefore this approximation will be valid for
$\varepsilon$ small, and in particular in the case where the frequency
is close to a rational. 

\section{Critical surface and scaling behavior}
\label{sec:cri}

\subsection{Renormalization trajectories}

First, we derive the general solution for the renormalization
trajectories. We compute the renormalization trajectories in the
variables $\u{x}=(x=\log f_1,y=\log \tilde{f_2})$.  We denote
$\gamma_1(\omega,k)=\log \rho_1(\omega,k)$, $\gamma_2(\omega)=\log
\rho_2(\omega)$, and 
$\u{\gamma}=(\gamma_1,\gamma_2)$.  The action of the renormalization 
operator ${\mathcal{R}}_a$ becomes~:
\begin{equation}
  \label{eq:rafor}
  \u{x}'=N_a\u{x}+\u{\gamma}(\omega,k),
\end{equation}
where
$$
N_a=\left(\begin{array}{cc} a & 1\\ 1 & 0\end{array}\right).
$$
By iteration, we obtain the general equation for a renormalization
trajectory~:
\begin{equation}
  \label{eq:traj}
  \u{x}_n=N_{a_{n-1}}\cdots N_{a_0} \u{x}_0
  +\u{\gamma}(\omega_{n-1},k_{n-1})+\sum_{l=0}^{n-2}
  N_{a_{n-1}}\cdots N_{a_{l+1}} \u{\gamma}(\omega_l,k_l),
\end{equation}
where
\begin{eqnarray*}
  && \omega_n=1/\omega_{n-1}-a_{n-1},\\
  && k_n=1/(k_{n-1}+a_{n-1}).
\end{eqnarray*}
We denote $p_n/q_n=[a_0,\ldots,a_{n-1}]$ the $n$th convergent of the
frequency $\omega$ (with $p_0=0$ and $q_0=1$). Then
$$
M_n\equiv N_{a_{n-1}}\cdots N_{a_0}=\left(\begin{array}{cc} q_n & p_n\\
    q_{n-1} & p_{n-1}\end{array} \right).
$$
The matrix $N_{a_{n-1}}\cdots N_{a_{l+1}}$ is equal to $M_n
M_{l+1}^{-1}$~:
$$
N_{a_{n-1}}\cdots N_{a_{l+1}}=(-1)^{l+1} \left(\begin{array}{cc}
    q_n p_l-p_n q_l & p_n q_{l+1}-q_n p_{l+1} \\
    q_{n-1} p_l-p_{n-1} q_l & p_{n-1} q_{l+1}-q_{n-1}
    p_{l+1}\end{array} \right).
$$

\subsection{Critical surface}

Following Ref.~\cite{mack88}, we determine the critical surface that
separates points for which $\u{x}_n \to -\infty$ like $-q_n$
and from those for which $\u{x}_n \to +\infty$ like $q_n$.  Thus the
critical surface is obtained by the condition $\u{x}_n/q_n \to 0$
when $n\to \infty$. The first component of Eq.~(\ref{eq:traj})
becomes~:
$$
\frac{x_n}{q_n}=x+\frac{p_n}{q_n}y +\sum_{l=0}^{n-2}(-1)^{l+1}
\left[\left(p_l-\frac{p_n}{q_n} q_l\right)
  \gamma_1(\omega_l,k_l)+\left(\frac{p_n}{q_n}q_{l+1}-p_{l+1}\right)
  \gamma_2(\omega_l)\right]+\frac{\gamma_1(\omega_{n-1},k_{n-1})}{q_n}.
$$
We use the fact that $p_n/q_n\to \omega$. Then we have~:
\begin{equation}
  \label{eq:eqsurf}
  x+\omega y +\sum_{l=0}^{\infty}
  \left[\beta_{l}\gamma_1(\omega_l,k_l)+\beta_{l+1}
  \gamma_2(\omega_l)\right] =0,
\end{equation}
where $\beta_l=(-1)^{l+1}(p_l-\omega q_l)$. We notice that the second
component of Eq.~(\ref{eq:traj}) gives the same
equation~(\ref{eq:eqsurf}) for the critical surface. Following
Ref.~\cite{mack88}, one can prove that the sum in
Eq.~(\ref{eq:eqsurf}) converges if and only if $\omega$ satisfies the
Bruno condition~:
$$
\sum_{l=0}^{\infty} \frac{\log q_{l+1}}{q_l} < + \infty.
$$

\subsection{Scaling law near $\omega$ rational}

In order to investigate the behavior of the critical surface near a
rational frequency $P/Q=[a_0,\ldots,a_{J-1}]$, we assume that $\omega$
has a large entry in its continued fraction expansion such that~:
$$
\omega^{(n)}=[a_0,\ldots,a_{J-1},n,a_{J+1},\ldots].
$$
We consider the behavior of Eq.~(\ref{eq:eqsurf}) when $n$ goes to
infinity. In the sum, we have two terms~: one referring to $\gamma_1$
and the other one to $\gamma_2$. The dominant term for the first sum
is obtained when $l=J$ and is equal to
$$
\beta_J \gamma_1(\omega_J,k_J)=\beta_J\log\left[ \frac{(k_J+1)\cdots
    (k_J+n)}{(1-\omega_J)^2\cdots(1-(n-1)\omega_J)^2}
  \frac{1+n/k_J}{2^n}\right].
$$
The remainder of the series is of order $O(\log n /n)$.
The continued fraction expansion of $\omega_J$ is equal to
$[n,a_{J+1},\ldots ]$. Then we have $\omega_J=1/n +O(1/n^2)$.
Furthermore, $k_J$ is independent of $n$ since its continued fraction
expansion is $[a_{J-1},\ldots, a_0,b_0,b_1,\ldots]$.  We use the
following asymptotic expansions~:
\begin{eqnarray*}
  && \sum_{l=1}^n \log(k_J+l)=n\log n -n +O(\log n),\\
  && -2\sum_{l=1}^{n-1} \log (1-l\omega_J)=2n+O(\log n).
\end{eqnarray*}
These expansions are obtained by using Stirling's formula. Concerning
the expansion of $\beta_J$, we use the formula~:
$$
\beta_J=\frac{\omega_J}{q_J+q_{J-1}\omega_J},
$$
which follows from the fact that $\omega_J=-(q_J\omega
-p_J)/(q_{J-1}\omega -p_{J-1})$.  Since $\omega_J=1/n+O(n^{-2})$ and
$q_J=Q$, we have~:
\begin{equation}
  \label{eq:beta}
  \beta_J= \frac{1}{nQ}+O(n^{-2}).
\end{equation}
Thus the expansion of the first sum is the following one~:
$$
\sum_{l=0}^{\infty}\beta_{l}\gamma_1(\omega_l,k_l)=\frac{1}{Q}(\log
n -\log 2+1) +O\left(\frac{\log n}{n}\right).
$$
We rewrite the second sum as~:
$$
-2\sum_{l=0}^{\infty} \beta_{l+1} \log (1-a_l\omega_l).
$$
We use the fact that $1-a_l\omega_l=\omega_l\omega_{l+1}$.  The
dominant terms of the second sum are the ones for $l=J-1$ and $l=J$
(terms containing $\log n$).
Since $\omega_J\approx 1/n$, these terms are of order $O(\log n
/n)$. \\
In summary, near a resonance $P/Q$, the asymptotic form of the
critical surface is~:
$$
x+\frac{P}{Q} y = -\frac{1}{Q}(\log n-\log 2 +1) +O\left(\frac{\log
    n}{n}\right).
$$
In order to express this behavior in terms of the distance from
$\omega$ to $P/Q$, we have from Eq.~(\ref{eq:beta})~:
$$
|\omega-P/Q|=\frac{1}{nQ^2}+O(n^{-2}).
$$
Then we have~:
$$
x+\frac{P}{Q} y \approx \frac{1}{Q}\log |\omega-P/Q|+c_Q.
$$
where $c_Q=(\log 2 -1+2\log Q)/Q$.  We consider the following
one-parameter family of Hamiltonians~(\ref{eq:hamrp}) with
$f_1=f_2=\varepsilon$ (denoted $F_2$ in Sec.~\ref{sec:equ}).  
We have the following scaling law for the
critical function of this one parameter family~:
\begin{equation}
  \log \varepsilon_c(\omega) \approx \frac{1}{P+Q} \log |\omega-P/Q|
  +c_{P,Q}(k),
\end{equation}
where $c_{P,Q}(k)=(\log 2-1+2\log Q-2P\log (kQ/P))/(P+Q)$ for $P\not=
0$, and $c_{0,1}=\log 2 -1$. We notice that near the main resonance
$0/1$ the characteristic exponent of the scaling law is equal to one
as it has been analytically found for the standard map in
Ref.~\cite{carl00}, but for a resonance $P/Q$ the scaling law is
different from the one of the standard map, and is characterized by a
characteristic exponent $1/(P+Q)$. Moreover we notice that the
exponent $\alpha_{P,Q}$ of the scaling law~(\ref{eq:sclaw}) is
independant of $k$ in that case, whereas $c_{P,Q}$ has a term in $\log
k$. In the integrable limit $k=0$, the constant $c_{P,Q}(k)$ and
consequently $\varepsilon_c(\omega)$ tend to infinity.\\
In summary, we have obtained a characteristic exponent
$\alpha_{P,Q}=1/(P+Q)$ of the scaling law~(\ref{eq:sclaw}) for the
one-parameter family $F_2$ by an
approximate renormalization transformation. 
In what follows, we analyze the scaling law~(\ref{eq:sclaw}) 
for other types of
one-parameter families of Hamiltonians which are linked to the family of
Hamiltonians $F_2$ by simple canonical transformations.

\section{Forced pendulum critical functions}
\label{sec:bru}

The critical function of the forced pendulum~(\ref{eq:edfp}) is linked
with the one of the one-parameter family of
Hamiltonians~(\ref{eq:gfp}) according to Eq.~(\ref{eq:numb})~:
\begin{equation}
  \label{eq:edfpH}
  \varepsilon_{fp}(1-\omega)=\varepsilon_{fp}(\omega)
  =\omega^2\varepsilon_{H}(1/\omega-1).
\end{equation}
Near a resonance $P/Q$ where $0<P<Q$ for Hamiltonians~(\ref{eq:edfp}),
the corresponding frequency for
$\varepsilon_H$ which is $1/\omega-1$, is near a
resonance $(Q-P)/P$ for Hamiltonians~(\ref{eq:monH}). Thus the scaling
law is the following one~:
\begin{equation}
  \label{eq:sca}
  \log\varepsilon_{fp}(\omega)\approx \frac{1}{Q}\log |\omega-P/Q|  
  +C_{P/Q},
\end{equation}
where
$$
C_{P/Q}=\frac{1}{Q}(\log 2 -1 +2\log Q)+2\frac{P}{Q}\log
\frac{P}{Q} +2\left(1-\frac{P}{Q}\right) \log
\left(1-\frac{P}{Q}\right).
$$
We notice that $C_{P/Q}$ is symmetric~: $C_{P/Q}=C_{1-P/Q}$.\\
The scaling law~(\ref{eq:sca}) has the same characteristic exponent
$1/Q$ as the scaling law found for the standard map by Carletti and
Laskar~\cite{carl00}. This exponent is consistent with the fact that
the standard map and the forced pendulum
Hamiltonian~(\ref{eq:edfp}) have the following symmetry for the
critical function~:
$\varepsilon_c(1-\omega)=\varepsilon_c(\omega)$. \\

The singular behavior of the critical function for the standard map 
and for the forced pendulum
model~(\ref{eq:edfp}) is linked to the singular behavior of the Bruno
function~\cite{marm97} $B : {\Bbb R}\setminus {\Bbb Q} \to {\Bbb R}^+ 
\cup\{\infty\}$ which is defined as follows~: Let $x$ denotes a real
number, $\langle x\rangle$ is its nearest integer (i.e.\ $\langle
x\rangle=[x+1/2]$, where $[\, \cdot \, ]$ is the integer part), 
and $\Vert x\Vert$ is the distance between $x$ and
$\langle x\rangle$. For an irrational frequency $\omega$, we consider the
sequences $\{\theta_k\}$ and $\{\beta_k\}$ defined by
\begin{eqnarray*}
&& \theta_k=\Vert \theta_{k-1}^{-1} \Vert,\\
&& \beta_{k-1}=\prod_{i=0}^{k-1} \theta_i,
\end{eqnarray*}
for $k \geq 1$ and $\theta_0=\Vert \omega\Vert$, and we define
$B(\omega)$ by
$$
B(\omega)=-\sum_{k=0}^{+\infty} \beta_{k-1} \log \theta_k,
$$
with $\beta_{-1}=1$.\\
For the standard map,
the following conjecture is supported by numerical calculations~\cite{marm92,carl00}~:\\
\indent \textit{${\mathcal{C}}$ : The function 
$\varepsilon_c(\omega)/\exp(-B(\omega))$ is a bounded function on
$[0,1]$.}\\
A strongest statement has been conjectured in Ref.~\cite{marm92},
by requiring that this function is continuous.
These conjectures come from the study of the radius of convergence of
the Lindstedt series of the standard map or complex analytic maps (see
for instance, Refs.~\cite{davi94,marm90,mous99}), and supported by
renormalization arguments~\cite{yocc95,marm97}. For the critical
function of the standard map, this problem is not solved even if
recent numerical results obtained in Ref.~\cite{carl00} 
suggest that the conjecture $\mathcal{C}$ holds.
For the approximate critical function we derive for the
forced pendulum~(\ref{eq:edfp}), we
expect that this conjecture is also true since it has the same
singular behavior near rationals as the one of the function
$\exp(-B(\omega))$.\\
However, the conjecture $\mathcal{C}$ is not satisfied in the
above statement, and this comes from the fact that one has to take
into account the right argument for the Bruno function, which is not
always $\omega$. In what follows, we analyze the critical function of
the forced pendulum~(\ref{eq:gfp}) in order to clarify the right choice  
of the argument, and to reformulate the conjecture $\mathcal{C}$ in 
a more appropriate form.\\

The critical function of the one-parameter family of
Hamiltonians~(\ref{eq:gfp}), denoted $\varepsilon_{fp}'(\omega)$, is linked to 
the critical function of the Hamiltonian~(\ref{eq:monH})
according to Eq.~(\ref{eq:numb})~:
$$
\varepsilon'_{fp}(\omega)=a^{-2}(d-c\omega)^2\varepsilon_{H}(\omega'),
$$
for $\omega\in [b/a,d/c]$, where
$\omega'=-(a\omega-b)/(c\omega-d)$. Moreover the critical function
$\varepsilon'_{fp}(\omega)$ is linked with the critical function
$\varepsilon_{fp}(\omega)$ by the following relation
 (see Eq.~(\ref{eq:edfpH}))~:
$$
\varepsilon_{fp}'(\omega)=a^{-2}((a-c)\omega+d-b)^2
\varepsilon_{fp}(\omega''),
$$
where 
\begin{equation}
  \label{eq:om''}
  \omega''=\frac{a\omega-b}{(a-c)\omega +d-b}.
\end{equation}
Near a rational frequency $P/Q$, the scaling law~(\ref{eq:sclaw}) is
satisfied with 
\begin{equation}
  \alpha_{P,Q}=1/((a-c)P+(d-b)Q),
\label{eq:chalp}
\end{equation}
since for the
one-parameter family of Hamiltonians~(\ref{eq:monH}), this exponent is
equal to the inverse of the
sum of the numerator and the denominator of the rational frequency. 
Thus the
singularities of these functions $\varepsilon_{fp}'$ are of a
different nature than the ones of the Bruno
function $B$. However, $\varepsilon'_{fp}(\omega)$ has the same
singularities as $\exp (-B(\omega''))$ where $\omega''$ is given by
Eq.~(\ref{eq:om''}). The conjecture
\textit{$\mathcal{C}$} has to be replaced by
the following one~:\\
\indent \textit{${\mathcal{C}}'$ : The function 
$\varepsilon_c(\omega)/\exp(-B(\omega''))$ is a bounded function on
$[b/a ,d/c]$}.\\
Practically the nature of the singularities is not changed, and the
scaling law~(\ref{eq:sclaw}) is general with the exponent $1/Q$ for a
typical one-parameter family of Hamiltonians~: The
two main resonances for the Hamiltonian~(\ref{eq:gfp}) are $b/a$
and $d/c$. They correspond to $\omega''=0$ and $\omega''=1$ which are
the two main resonances of the forced pendulum~(\ref{eq:edfp}). The
characteristic exponent $\alpha_{P,Q}$ close to these main resonances
is equal to 1, since $(a-c)P+(d-b)Q=1$ for $P/Q=a/b$ or $d/c$. Thus
the characteristic exponent of the scaling law~(\ref{eq:sclaw}) close
to the main resonances is
equal to one for all the Hamiltonians~(\ref{eq:gfp}) even if the main
resonances are not at the same place as the ones of the forced pendulum
model~(\ref{eq:edfp}). 
Moreover, the characteristic exponent of the secondary
resonances is equal to $1/Q$ where $Q$ if the order of the resonance
in the parameter $\varepsilon$ in the perturbation expansion. 
For instance, the resonance
$(d+b)/(a+c)$ is of order $\varepsilon^2$. Its characteristic exponent
is equal to $1/2$ since $(a-c)P+(d-b)Q=2$ for $P/Q=(d+b)/(a+c)$.

\ack

We acknowledge useful discussions with G Benfatto, T Carletti, G 
Gallavotti, H R Jauslin, H Koch, J Laskar, R S MacKay, S
Marmi and J C Yoccoz. CC thanks support from the Carnot Foundation.

\end{document}